\documentclass[%
 aip,
 amsmath,amssymb,
 reprint,%
]{revtex4-1}

\usepackage{graphicx}
\usepackage{dcolumn}
\usepackage[version=4]{mhchem}
\usepackage{bm}
\usepackage[dvipsnames]{xcolor}
\usepackage{soul}
\usepackage[utf8]{inputenc}
\usepackage[T1]{fontenc}
\usepackage{mathptmx}
\usepackage{parskip}
\usepackage{placeins}
\usepackage{siunitx}
\begin{document}
\setlength{\parskip}{0pt}
\setlength{\parindent}{2em}
\preprint{AIP/123-QED}

\title[Semi-insulating 4H-SiC lateral bulk acoustic wave resonators]{Semi-insulating 4H-SiC lateral bulk acoustic wave resonators}

\author{B. Jiang}
 \email{jiang561@purdue.edu.}
\author{N. Opondo}%

\author{S. Bhave}
 \affiliation{%
OxideMEMS Lab, Purdue University, West Lafayette, Indiana 47907, USA
}%

\date{\today}

\begin{abstract}
Silicon carbide (SiC) excels in its outstanding mechanical properties, which are widely studied in Microelectromechanical systems (MEMS). Recently, the mechanical tuning of color centers in 4H-SiC has been demonstrated, broadening its application in quantum spintronics. The strain generated in a mechanical resonator can be used to manipulate the quantum state of the color center qubit. This work reports a lateral overtone mechanical resonator fabricated from a semi-insulating (SI) bulk 4H-SiC wafer. An aluminum nitride (AlN) piezoelectric transducer on SiC is used to drive the resonance. The resonator shows a series of modes with quality factors (Q) above 3000. An acoustic reflector positioned at the anchor shows a 22\% improvement in the Q at 300 MHz resonance and suppresses the overtone modes away from it. This monolithic SiC resonator allows optical access to the SiC color centers from both sides of the wafer, enabling convenient setup in quantum measurements.
\end{abstract}

\maketitle


Silicon carbide (SiC) is known for its hardness, elasticity at high temperatures, high thermal conductivity, and chemical inertness, making it suitable for the next generation MEMS components. Recently, the color centers in SiC, such as divacancies and silicon vacancies, have been demonstrated as viable platforms for quantum sensing and computation \cite{Son2020DevelopingSpintronics}. In particular, researchers have been able to use surface acoustic wave (SAW) in SiC resonators to control the spin states inside the acoustic cavity \cite{Whiteley2019SpinphononAcoustics}. MEMS researchers have usually used SiC-on-Insulator (SiCOI) technology, which involves epitaxially grown SiC on an oxide layer. For example, the 3C-SiC Lateral Overtone Bulk Acoustic Wave Resonator (LOBAR) devices developed by Gong et al.\cite{6127890} show outstanding resonance performance and are easy to fabricate at wafer scale. However, SI 4H-SiC, a polytype that cannot be grown epitaxially, stands out as a unique candidate for its optical properties \cite{Falk2013PolytypeCarbideb} and its potential as the host for single quantum emitters \cite{Castelletto2014ASourceb}.

Recently, Lekavicius et al. introduced a diamond lamb wave resonator \cite{Lekavicius2019DiamondCenters} and created bright NV centers in it, a similar system to color centers in SiC. However, their devices are not integrated with a piezoelectric transducer, and therefore cannot be mechanically actuated. We have previously fabricated cantilever resonators made from SI 4H-SiC \cite{8808560} resonating near 10 MHz and showed that the process preserves the fluorescent quality of the divacancies. The color center spin's mechanical Rabi frequency increases linearly with the frequency of the mechanical resonance \cite{MacQuarrie:15}, so higher-frequency resonators are desired for fast quantum control. In this work, we report the development of 300 MHz acoustic wave resonators actuated by AlN piezoelectric layer on our SiC platform.

\begin{figure}[htbp]
\centering
\includegraphics[width=0.45\textwidth]{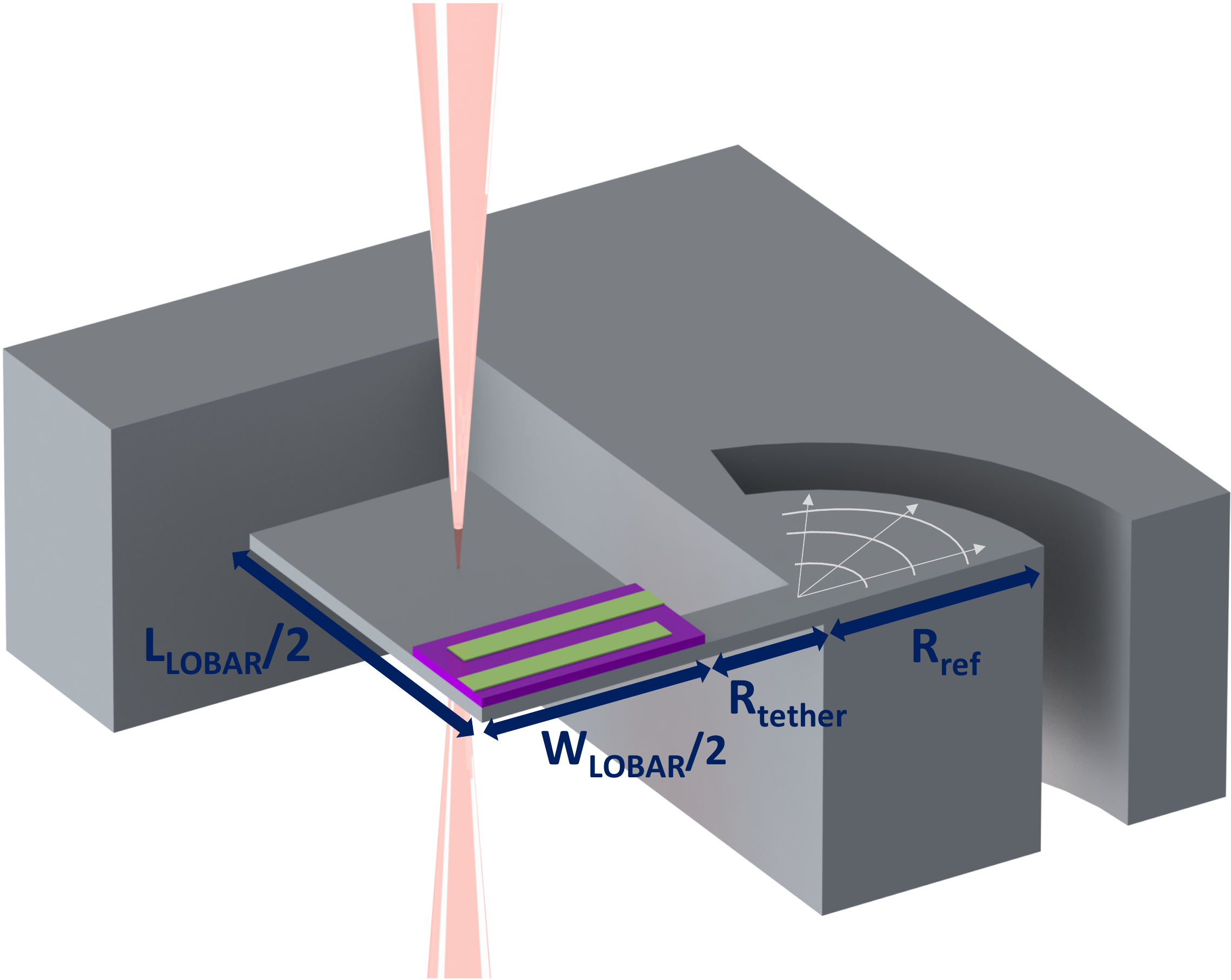}
\caption{\label{fig:design} 1/4th of the LOBAR resonator connected to the anchor area. The white lines indicate the mechanical wave leaking from the anchor, met by the reflector. The laser used for quantum measurement can be applied from both the upper and the lower surface of the LOBAR's extending wings.}
\end{figure}

A 3D illustration for the device we made is shown in FIG. \ref{fig:design}. We design the LOBAR devices based on a \SI{200}{\micro\metre} thick 4H-SiC wafer. The LOBAR has an AlN transducer in the center, which has electrically alternating inter digitated transducer (IDT) fingers generating a laterally vibrating acoustic wave. The LOBAR generates overtones modes in the frequency spectrum, because of the SiC wings extending beyond the IDT-covered area on both sides. We also place an acoustic reflector \cite{Harrington_2011}, an arc-shaped trench through the bulk wafer, centering at the LOBAR's anchor point. The leaking acoustic wave from the anchor incident on the trench will be reflected back into the resonator body, thus reducing the acoustic energy loss. For a reflector positioned at $R_{ref}$ from the tether, the total wave propagating length is $2R_{ref}$. When $R_{ref}=\lambda /2 + n \cdot \lambda  (n=0,1,2,\ldots$ and $\lambda$ is the acoustic wavelength), the highest Qs are achieved due to interference.

For quantum applications, when a laser is introduced to excite the color center spins in the SiC, our device allows laser access to both surfaces of the SiC LOBAR, illustrated in FIG. \ref{fig:design}. Compared with other MEMS resonators on SiC and diamond \cite{Whiteley2019SpinphononAcoustics, MacQuarrie:15,Maity2020CoherentDiamondb, Chen2019EngineeringResonator}, which are unreleased and allow only optical access from one side, our resonators would extend the sensing flexibility of spins. Optical access on one side will allow the sensing proximity to a measurand on the other, which is important in nanoscale sensing \cite{Thiel973}.

\begin{figure}[htbp]
\centering
\includegraphics[width=0.48\textwidth]{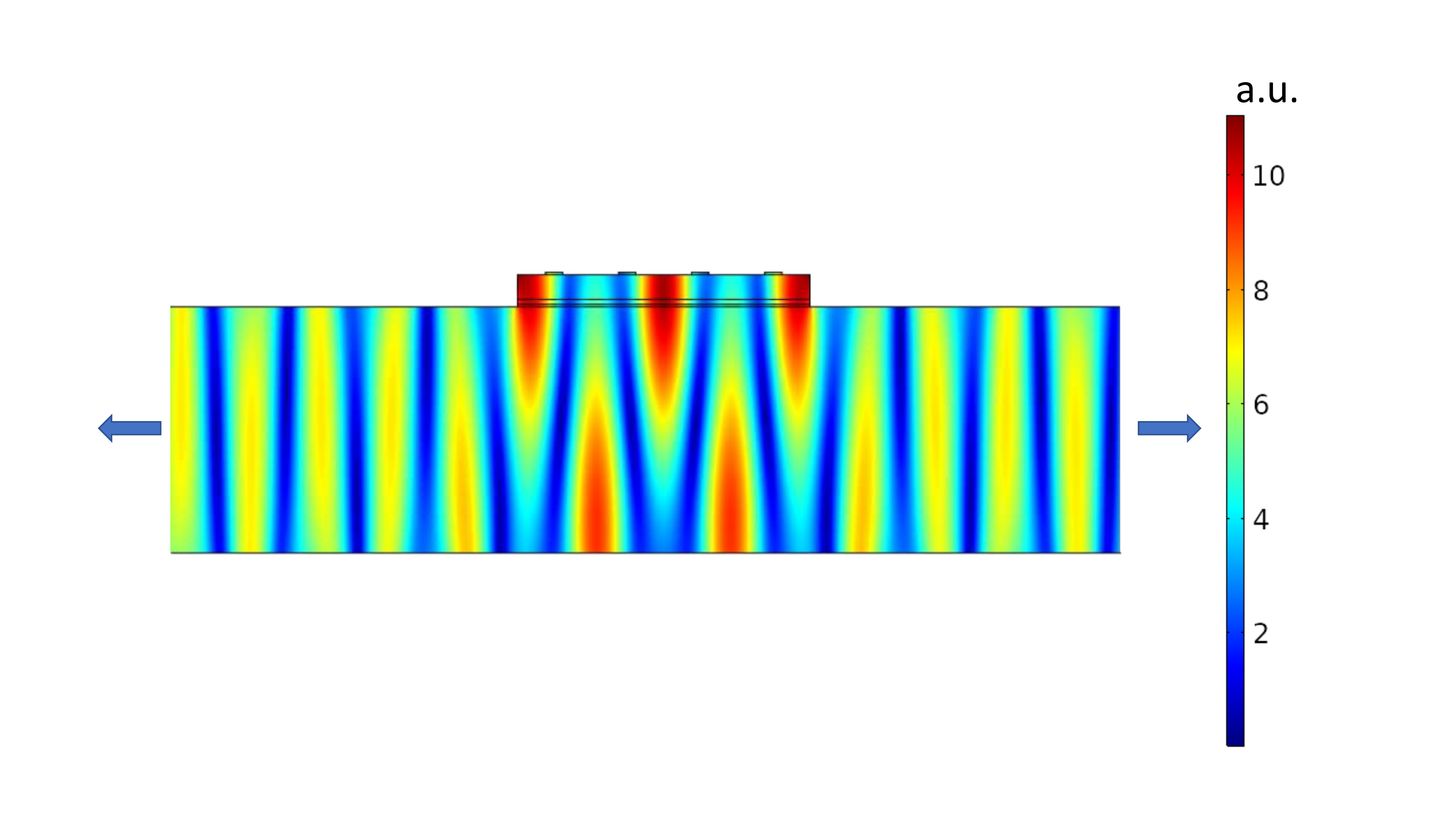}
\caption{\label{fig:comsol} Mode shape of the LOBAR device, shown as the displacement in COMSOL simulation. The IDT fingers in the center region create the alternative mechanical vibration to both sides.}
\end{figure}

To find the appropriate design parameters, we performed a 2D COMSOL Multiphysics modal analysis based on the design shown in FIG. \ref{fig:comsol}. Because of the simulation and fabrication limitations, we set the LOBAR length as \SI{1600}{\micro\metre}, width as \SI{300}{\micro\metre}, and the thickness as \SI{10}{\micro\metre}, while $R_{tether}$ as \SI{90}{\micro\metre} and $R_{ref}$ as \SI{60}{\micro\metre}. The distance between 2 IDT fingers is \SI{20}{\micro\metre}, leading to a resonance frequency around 300 MHz.

\begin{figure}[htbp]
\centering
\includegraphics[width=0.48\textwidth]{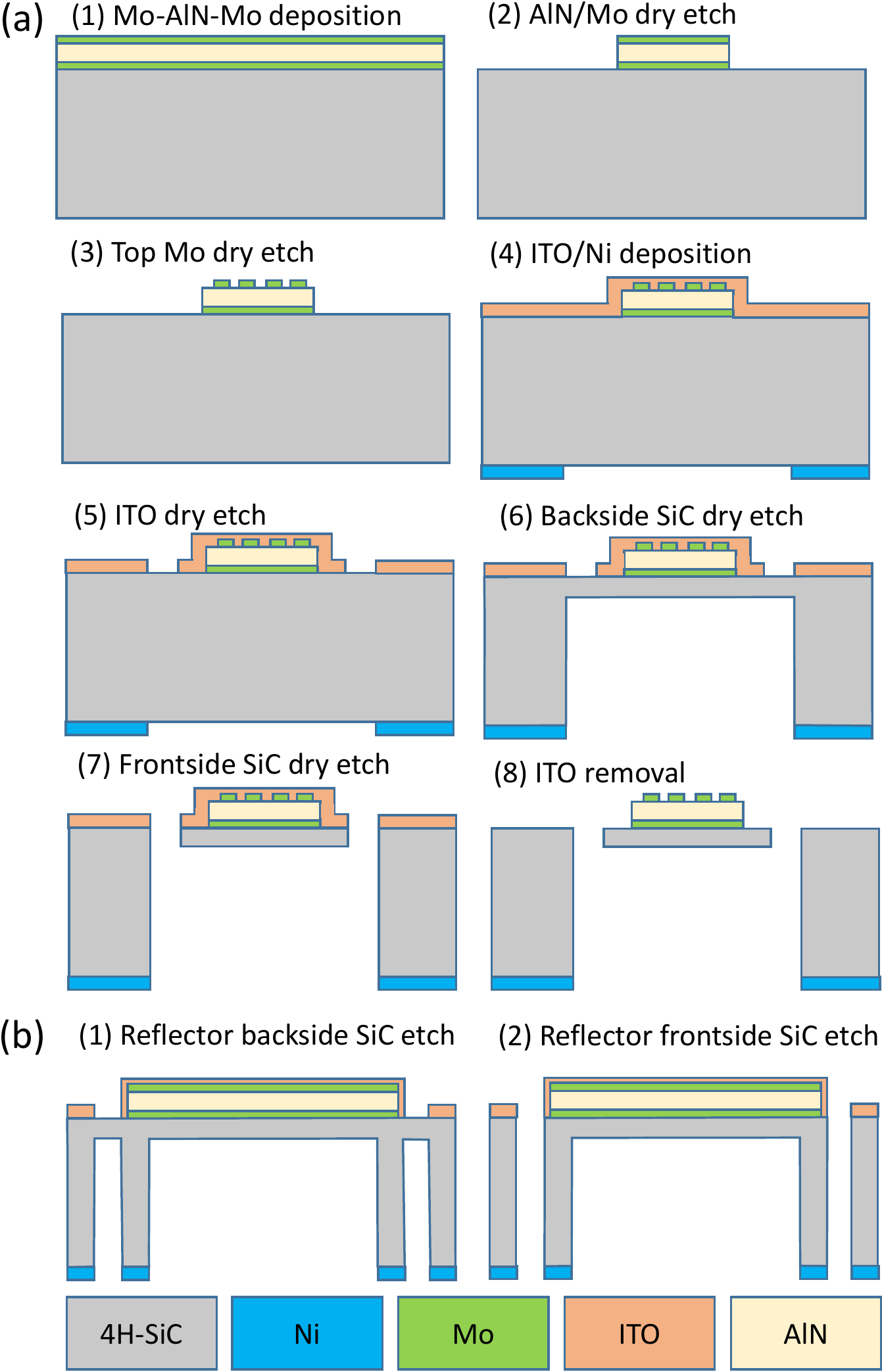}
\caption{\label{fig:flow} (a) Fabrication process of the SiC LOBAR resonator. The cross-section is along the longitudinal direction of the LOBAR. (b) Fabrication of the reflectors. The figures correspond to step (6) and (7) in (a). The cross-section is along the transverse direction of the LOBAR. Compared with the previous acoustic reflectors \cite{Harrington_2011,8597472} which were only defined by frontside etch, our front-backside etch allows a reflector through the whole wafer thickness. It may be used in future developments in anchor loss reduction.}
\end{figure}

We fabricate the SiC LOBAR devices from a \SI{500}{\micro\metre} thick SI 4H-SiC wafer purchased from Cree Inc, with resistivity larger than \SI{1E6}{\ohm . \cm}. The wafer is thinned down to \SI{200}{\micro\metre} by chemical mechanical polishing (CMP). The detailed process after thinning is shown in FIG. \ref{fig:flow}(a): (1)-(3) a piezoelectric stack consisting of \SI{1}{\micro\metre} AlN sandwiched between 100 nm Molybdenum (Mo) layer is deposited by OEM Group Inc. Optical lithography and dry etch are performed to define the piezoelectric transducers. (4) The key fabrication step involves a dual-mask followed by front-backside etch, enabling a wafer-scale dry release of the LOBARs. For the hardmask of the SiC etch, we deposit \SI{4}{\micro\metre} thick Nickel (Ni) on the backside \cite{8808560}. At the same time, we use electron beam evaporation to form an Indium Tin Oxide (ITO) thin layer (600 nm) as the etch mask for the frontside of SiC, since it can be dissolved by wet etch without damaging the metal electrodes. (5) The ITO layer is patterned by photoresist, then etched by \ce{Cl2} based plasma, to define the release window for LOBARs. (6) From the backside, the SiC is dry etched with \ce{SF6} and Ar plasma. The etch rate is about \SI{15}{\micro\metre} per hour. The etch is stopped when it reaches \SI{190}{\micro\metre}, leaving a \SI{10}{\micro\metre} SiC membrane. (7) Turning over the wafer, we perform another dry etch on the frontside for \SI{10}{\micro\metre} until the LOBAR device is released. (8) Finally, we dissolve the ITO with 1:5=\ce{HCl}:\ce{H2O} diluted hydrochloric acid and leave the backside Ni mask on as it does not affect our measurement. The acid does not attack the AlN transducer stack. Concurrently, a through wafer \SI{200}{\micro\metre} acoustic reflector is defined using the frontside \SI{10}{\micro\metre} and \SI{190}{\micro\metre} deep etches, shown in FIG. \ref{fig:flow}(b).


\begin{figure}[htbp]
\centering
\includegraphics[width=0.48\textwidth]{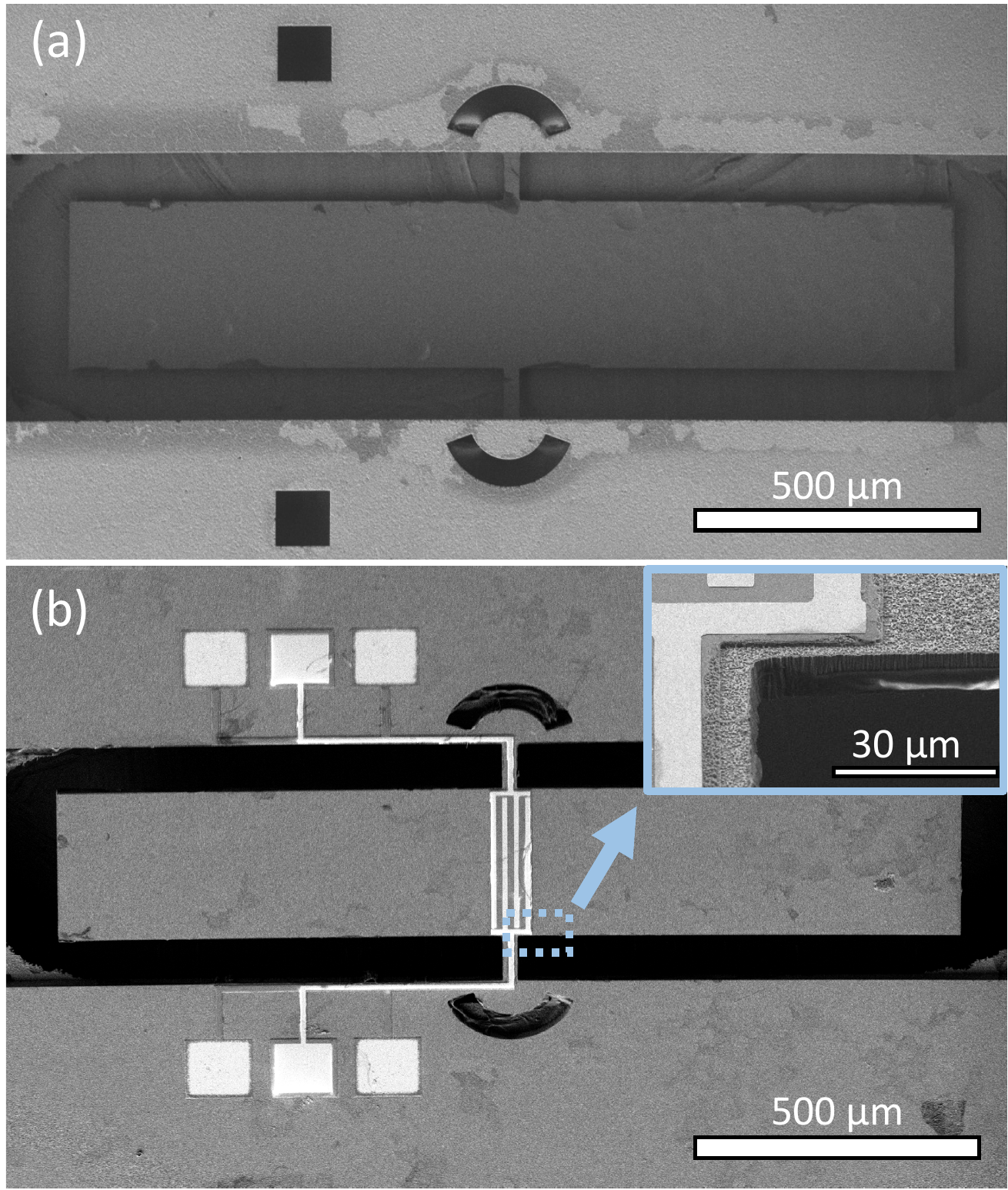}
\caption{\label{fig:sem} (a) Backside of the LOBAR. The reflectors go through the whole wafer and have openings on both sides of the wafer. The LOBAR is \SI{1600}{\micro\metre} long and \SI{300}{\micro\metre} wide. (b) Frontside of a released LOBAR. The enlarged picture shows SiC surface damage during the AlN etch.}
\end{figure}

The final devices are inspected with Hitachi S-4800 Field Emission SEM. The images are shown in FIG. \ref{fig:sem}. The SiC LOBAR is released using our front-backside etch. One challenge in the fabrication is that the main LOBAR release window is much larger than the reflector, resulting in different etch rates in these two regions. When the backside etch reaches \SI{190}{\micro\metre} in the LOBAR area, the reflector etch on the backside is shallower. We overcome the issue by an over-etch on the frontside with an ITO hardmask. ITO shows selectivity of 1:50 over SiC, and survives the extensive etch on the frontside. FIG. \ref{fig:sem}(b) shows the piezoelectric transducer defined on top of SiC LOBAR. Even though the dry etch of AlN with \ce{BCl3} stops on the SiC layer, it leaves a rough surface due to heavy \ce{Cl^-} ion bombardment.

\begin{figure}[htbp]
\centering
\includegraphics[width=0.48\textwidth]{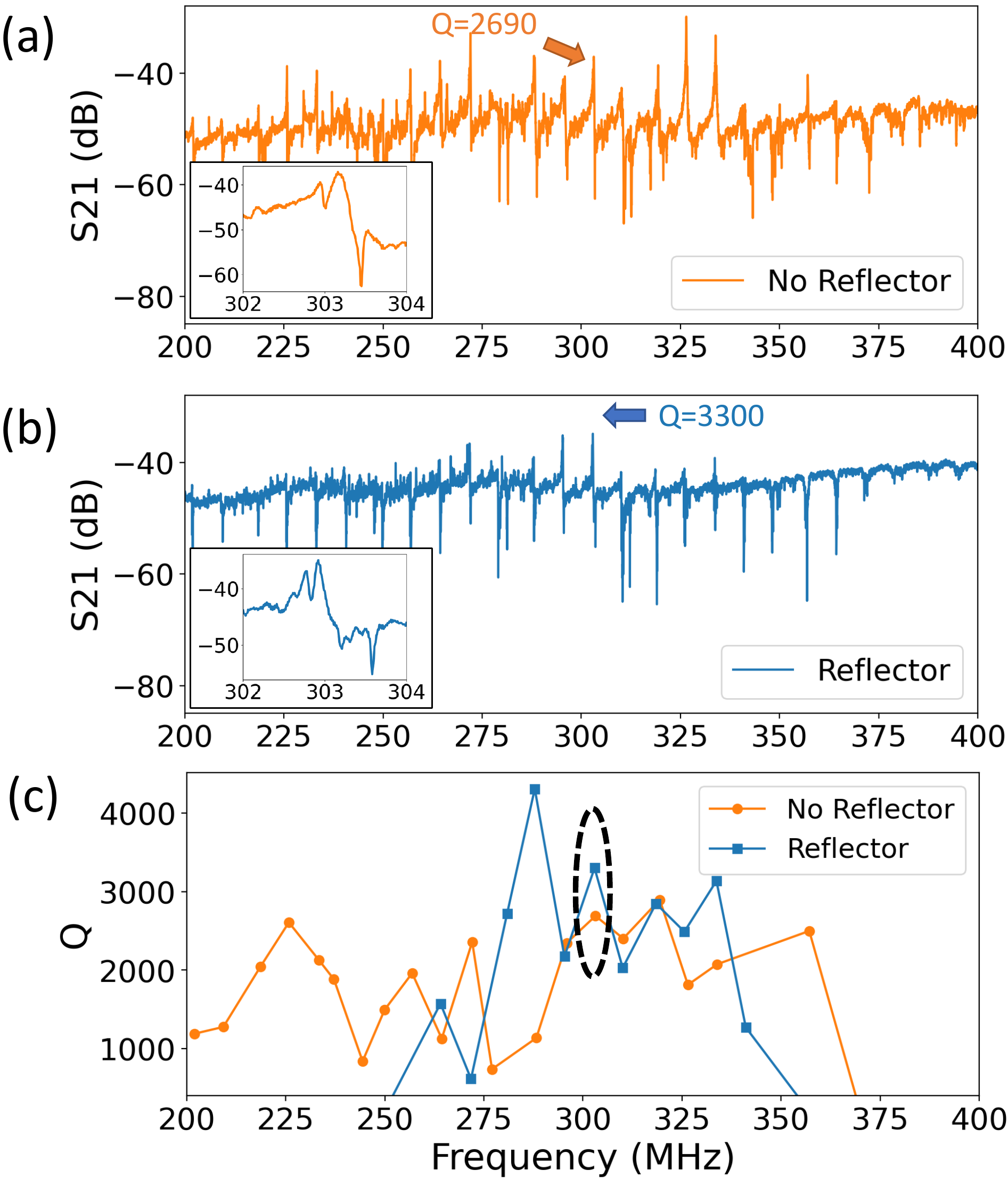}
\caption{\label{fig:measurement} 2-port transmission measurements of: (a) LOBAR without the reflector. (b) LOBAR with the reflector, $R_{ref}=1.5\lambda$. The reflector is frequency-sensitive, suppressing the resonances away from 300 MHz. (c) Q measured in both devices' overtone modes. The circled two modes are the marked ones and shown in figure insets in (a) and (b) respectively.}
\end{figure}

The resonators are characterized on a Cascade Microtech 11971B Probe Station with a Keysight N5230A PNA-L Network Analyzer and 2 Cascade Microtech ACP40 GSG probes at room temperature and under atmosphere conditions. In FIG. \ref{fig:measurement}, we compare the transmission measurements of LOBARs with and without the reflectors. The Qs are calculated with the Lorentzian peak fitting method. The piezoelectric transducer couples to multiple overtones in the resonator. We define the target mode of the LOBAR as the mode with maximum coupling with the transducer. We realize a 22\% improvement on Q in the target LOBAR mode with the reflector. In FIG. \ref{fig:measurement}(c), the LOBAR with the reflector has increased Qs around 300 MHz, while the Qs of other overtone modes quickly decline and vanish below the measurement noise floor outside a 100 MHz wide range. Meanwhile, the LOBAR without the reflector displays lower Qs across a wider frequency spectrum.

As it is discussed in FIG. \ref{fig:design}, for a designed $R_{ref}$=\SI{60}{\micro\metre}, the Q is maximized when the wavelength $\lambda$=\SI{40}{\micro\metre} and n=1. With an acoustic velocity of 12500 m/s \cite{Chinone1989ApplicationsII}, this wavelength corresponds to a frequency of 300 MHz in 4H SiC. Other modes are affected by the miss-matched interference and suffer a Q loss. A LOBAR has multiple modes with similar Qs and piezoelectric coupling. Appropriate design and positioning of the acoustic reflector enhance the target mode and attenuate the adjacent modes' Qs. This Q boost will enhance the strain-spin coupling in the color center applications in SiC. Our design will also simplify spur-free RF LOBAR oscillators \cite{Kourani2020AResonator}.


The SI 4H-SiC resonators’ Qs are similar to the previous 3C-SiC LOBARs \cite{6127890} and the SCHBAR in diamond \cite{Chen2019EngineeringResonator}, but still lower than the LOBARs made by Ziaei-Moayyed et al.\cite{Ziaei-Moayyed2011}. Future fabrication improvement to increase the Q will include reducing the surface roughness in AlN etch on the frontside and reducing the dimples, and improving the etch depth uniformity on backside and increasing the sidewall smoothness.

In conclusion, we have introduced a monolithic high-frequency lateral bulk acoustic wave resonator based on SI 4H-SiC, and improved its mechanical performance by a reflector design. Front-backside etch process provides unique flexibility to produce other types of resonators \cite{1174134, Hamelin2019MonocrystallineModesb,Gokhale2020EpitaxialAcoustodynamicsb,PRXQuantum.1.020102}. In the future, we shall explore the mechanical coupling to the color centers using our devices, and study the effect of a laterally vibrating acoustic wave on the quantum system. Furthermore, since no other substrate material is needed and the optical accessibility is on both sides, we can combine this process with the development of 2D material on SiC \cite{Borysiuk2014StructuralStudies} and study the acoustic effect on the interface between the two materials.


\vspace{2ex}
This work was conducted at Birck Nanotechnology Center in Purdue University. The research was supported by NSF RAISE-TAQS Award \#1839164. The CMP was completed at NOVASiC, and the AlN deposition was performed by Plasma-Therm.

\vspace{2ex}
The data that support the findings of this study are openly available in Zenodo at \url{https://doi.org/10.5281/zenodo.4568410}.

\bibliography{references}

\end{document}